\documentclass{aa}
\usepackage{amsmath}

\usepackage[svgnames]{xcolor}
\definecolor{blue}{RGB}{0,0,255}
\usepackage{caption}
\usepackage[font={color=blue},figurename=Fig.,labelfont={it}]{caption}
\usepackage{bm}
\usepackage{subcaption}
\usepackage[varg]{txfonts}
\usepackage{graphicx}
\usepackage{natbib}
\usepackage{pdfpages}
\usepackage{wasysym}
\usepackage{rotating}
\bibpunct{(}{)}{;}{a}{}{,} % to follow the A&A style
\usepackage{hyperref}

\hypersetup{
    colorlinks=true,                          
    linkcolor=blue, 
    citecolor=blue, 
    urlcolor=blue,
    linktoc=all
} 

\newcommand{\fig}{\textcolor{blue}{\textit{Fig.}}}

\begin{document}
\title{On two formulations of polar motion and identification of its sources}
\author{F. Lopes\inst{1}
\and V. Courtillot\inst{1}
\and J-L. Le Mouël\inst{1}
\and D. Gibert\inst{4}
}
\institute{{ Universit\'e de Paris, Institut de Physique du globe de Paris, CNRS UMR 7154, F-75005 Paris, France}
\and { LGL-TPE - Laboratoire de G\'eologie de Lyon - Terre, Planètes, Environnement, Lyon, France}
}
\date{}

\abstract {Differences in interpretation may arise from differences in formulation of the equations of celestial mechanics. This paper focuses on the Liouville-Euler system of differential equations. In a “modern” presentation of the equations, variations in polar motion and variations in length of day are decoupled. Their source terms (or excitation functions) result from redistribution of masses and torques. In the “classical” presentation, polar motion is governed by the inclination of Earth’s rotation pole and the derivative of its declination (close to the derivative of length of day - lod). These are coupled by the Liouville-Euler system. In the “classical” approach, all “source” terms are astronomical. The Liouville-Euler system allows one to determine the period of the Euler free oscillation (theoretical period: 306 days). The observed period (actually a doublet, known as the Chandler free oscillation; ~430 and ~433 days) is much longer. It varies with polar inclination from 306 to 578 days. Moreover, its envelope is strongly modulated, reaching a quasi-minimum around 1930 (with a $\pi$ phase jump). The transition of the double period took place at the time of the 1930 phase jump. The duration and modulation of the Chandler wobble require a source of excitation. Elasticity of the Earth, large earthquakes, or external forcing by the fluid envelopes have been successively invoked in the “modern” approach. In the “classical” approach, the duration and modulation of the Chandler wobble can simply be accounted for by variations in polar inclination. The “classical” approach also implies that there should be a link between the rotations and the torques exerted by the planets of the solar system. There is remarkable agreement between the sum of forces exerted by the four Jovian planets and components of Earth’s polar motion. Since 1970, the length of day tends to decrease. In the “modern” approach, motions in the fluid envelopes have been proposed as potential causes of this decrease. A recent acceleration of rotation velocity, that contradicts previous models, finds a simple explanation with the “classical” formalism. A sufficiently strong source of energy can be found: components of motions with luni-solar periods account for 95\% of the total variance of the lod signal. Analysis of lod (using more than 50 years of data) finds nine components, all with physical sense: first comes a “trend”, then pseudo-oscillations with periods ~80 yr (Gleissberg cycle), 18.6 yr (Saros cycle), 11 yr (Schwabe cycle), 1 year and 0.5 yr (Earth revolution and first harmonic), 27.54 days, 13.66 days, 13.63 days and 9.13 days (Moon synodic period and harmonics). The lod contains a richer collection of high frequency components than polar motion that can be seen as a consequence of the derivative operator. The longer periods, 1 yr, 11 yr, 18.6 yr and ~80 yr are common to lod and polar motion. The Euler-Liouville system of equations in the “classical” analysis implies a strong link between the two: the straightening of the inclination of the axis of rotation should (and does) accompany a decrease in lod Also, if this link is valid, all the components with extraterrestrial periods should be present in the series of sunspots and indeed they are.}

\titlerunning{On two formulations of polar motion}
\maketitle

\section{Introduction} 
	Over more than two centuries, scientists have attempted to measure and to explain the variations in the length of the day (lod), or the equivalent rotation velocity of Earth, and changes in the geographical location of the pole of rotation, that is the place where the rotation axis intersects the Earth surface. A thorough treatment is in the \textit{Treatise of celestial mechanics} of Pierre-Simon de Laplace  (1749-1827; \cite{Laplace1799}) where the great scientist derives the system of differential equations that fully describes the motions of the rotation axis of any celestial body, among others Earth. This system has come to be known as Liouville-Euler after mathematicians Leonhard Euler (1707-1783; the first to establish this system for pole motion) and Joseph Liouville (1809-1882; he complemented the theory of partial differential equations, showing in a theorem that the volume of phase space of a system is constant along trajectories of the system – in modern terms). The theory has been confirmed and elaborated on by a number of authors,  \citet{Poincare1899} among them. Recent formulations are found in many papers and textbooks. In the present note, we focus on that of  \citet{Lambeck2005}. In a first section, we recall the theoretical derivations of Laplace and Lambeck and show aspects in which they are formally identical, but also differences that can be significant and need to be explained. The main point is the identification of the sources (or excitation functions) of polar motion and length of day (lod). In a second section, we illustrate these differences by applying the theory to modern data of polar motion and lod. We discuss them and draw some conclusions in the final section.

\section{Two formulations of the Liouville-Euler equations}
\begin{figure}
\centerline{\includegraphics[width=\columnwidth]{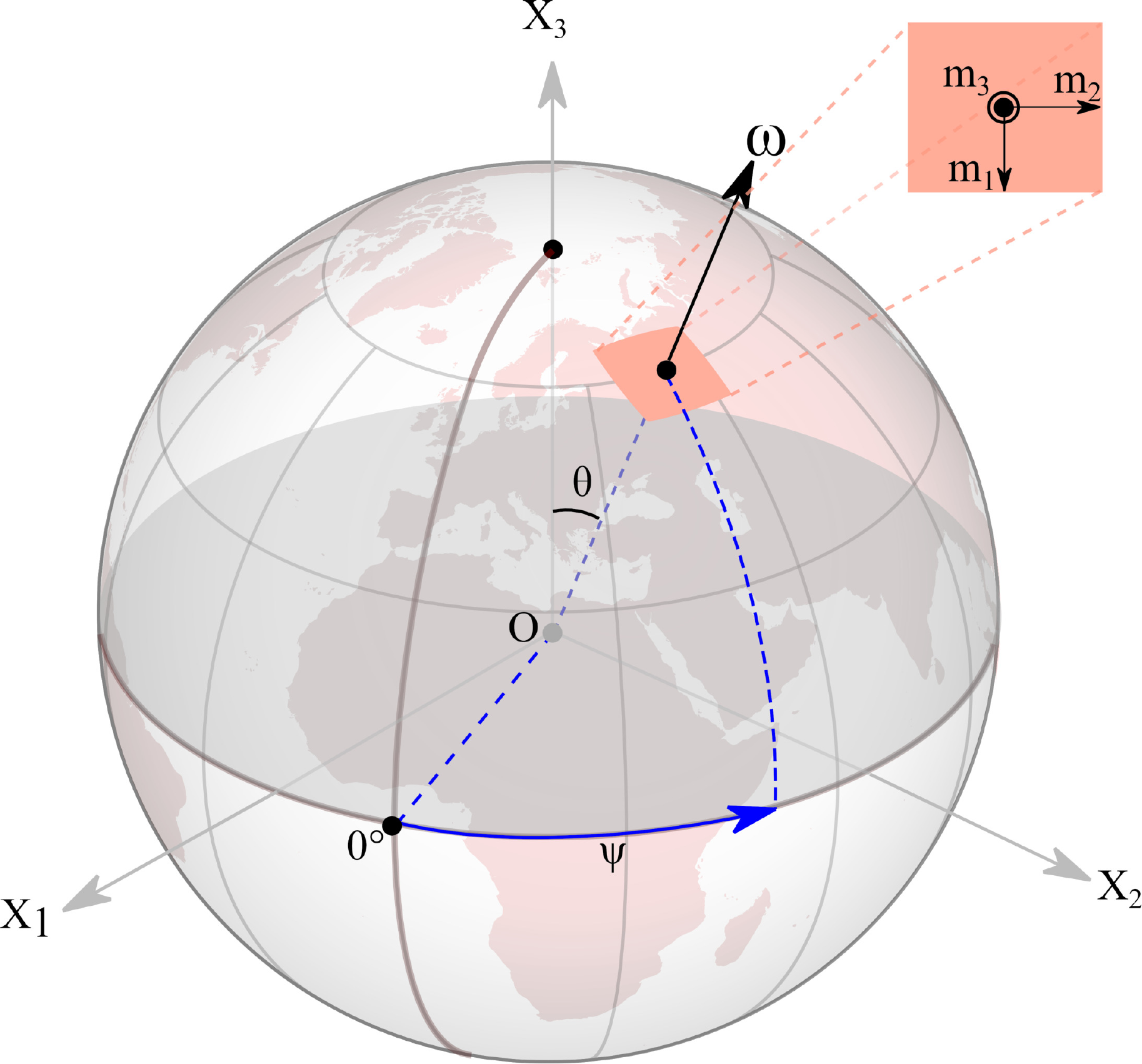}} 	
\caption{Terrestrial reference frame. $m_1$ and $m_2$ are the coordinates of the rotation pole. $\psi$  and $\theta$  are the declination and inclination introduced by \citet{Laplace1799}. }
\label{Fig:01}
\end{figure}

\citet{Lopes2021} and \citet{Courtillot2021} have recently recalled in some detail how \citet{Laplace1799}  derived the system of differential equations that was later to be named Liouville-Euler. We call “full” polar motion the vector consisting of the two spherical coordinates of the pole $m_1$ and $m_2$ (according to the $X_1 O X_2$ plane) and the third coordinate $m_3$, linked to the length of the day (\textit{cf.} \fig \ref{Fig:01}). The motion of the Earth’s rotation axis ($\omega$) can be seen as the combination of three Euler angles $\omega_1$, $\omega_2$ and $\omega_3$. The rotation axis moves only very small distances from its mean position (at least over the past century of continuous measurements) and one can write:
\begin{equation}
	\begin{aligned}
		\omega_1 &= \Omega.\textrm{m}_1 \\
		\omega_2 &= \Omega.\textrm{m}_2 \\
		\omega_3 &= \Omega.(1+\textrm{m}_3) 
	\end{aligned}
	\label{eq:01}
\end{equation}

where $\Omega$ (=7.292115*10$^{-5}$ rad/s) is the Earth’s mean rotation velocity today computed on the last 3 decades. Applying the theorem of kinetic momentum to the rotation of a non rigid body and following \citet{Lambeck2005}, chapter 3, equations (\ref{eq:01}) lead to the set of Liouville - Euler equations (system 3.2.9 in \citet{Lambeck2005}):
\begin{equation}
	\begin{aligned}
		i (\dfrac{\dot{\textbf{m}}}{\sigma_r}) + \textbf{m}  & = \textbf{f} \\
		\dot{m}_3 &= f_3
	\end{aligned}
	\label{eq:02}
\end{equation}

where $i=\sqrt{-1}$, $\textbf{m} = m_1+im_2$, $\sigma_r$  is the Euler frequency ($=\dfrac{C-A}{A}\Omega$), $\textbf{f}$ (=$f_1+if_2$) and $f_3$ are the so-called excitation functions (\textit{e.g.} \citet{Lambeck2005}, chapter 4). $C$  and $A$ are respectively, the axial and equatorial moments of inertia of Earth (8.0365*10$^37$ kg.m$^2$ and 8.010*10$^37$ kg.m$^2$, see \citet{Chen2010}). In this derivation, the behavior of the pole position ($m_1$,$m_2$) and $m_3$ have been fully separated. Through $\sigma_r$ , ($m_1$,$m_2$) involve the (internal) terrestrial data $C$ and $A$. \citet{Lambeck2005}, page 34, writes: “\textit{$m_1$ and $m_2$ are the components of the polar motion or wobble and  $\Omega \dfrac{d \omega_3}{dt}$ is nearly the acceleration in diurnal rotation}”. The generally accepted reading (physical interpretation) of this formulation is that polar motion ($m_1$, $m_2$) is linked to geophysical excitation such as atmospheric or oceanic circulation, lithospheric and mantle convection or electromagnetic coupling, and that the $m_3$ component is linked to astronomical phenomena such as tides. \citet{Lambeck2005}, page 36) concludes: “ \textit{Equations (3.2.6) clearly separate the astronomical and geophysical problems.}”\\

This is one reason for which on long time scales mechanical properties of the mantle are called upon. Some of these excitations can vary along with climate variations. Such is the case in the theory of isostasy (e.g. \citet{Peltier1976}; \citet{Nakiboglu1980}). It is also the reason why one calls upon the ephemerids of the Moon and Sun in order to compute Earth tides (\textit{e.g.} \citet{Melchior1966}; \citet{Ray2014}) or to evaluate their influence on lod variations (e.g. \citet{Wahr1981}; \citet{LeMouel2019a}). \\

In the classical problem of the motion of a Lagrange top (\textit{e.g.} \citet{Perez1997}), the weight of the top plays the role of a perturbation of the full motion of the rotation axis. This is replaced by the astronomical torques of the Moon and Sun as excitations of the Earth’s rotation axis. This formalism has allowed one to compute the period of equinoctial precession (26,000 yr). This is how one is led to the Milankovitch theory of climate variations (\citet{Milankovic1920}). The role of other bodies in the solar system, mainly the large and remote Jovian (ice) planets, must be taken into account (\textit{e.g.} \citet{Laskar2004}; \citet{Laskar2011}), for they excite responses in eccentricity at very long periods (\textit{e.g.} 400 kyr). As is the case for the top, changes in Earth rotation perturb its revolution about the Sun. \\

Laplace’s “classical” formulation of the theory is formally identical to the “modern” formulation recalled above (\citet{Lambeck2005}). But it leads to a different interpretation (reading). \citet{Laplace1799}, page 74, system D,  deduces from the fundamental law of dynamics a system of equations that can be shown to be identical to (\ref{eq:02}). It reads (\citet{Laplace1799}, page 74, system D):
\begin{subequations}
\begin{equation}
	\begin{aligned}
		\textrm{d}p^{'} + \dfrac{B-A}{AB}.q^{'} r^{'}  \textrm{dt}&= \textrm{dN}.\cos(\theta)- \textrm{dN}^{'} .	\sin(\theta)\\
		\textrm{d}q^{'} + \dfrac{C-B}{CB}.r^{'}p^{'} \textrm{dt}&= -\{\textrm{dN}.\sin(\theta)+ \textrm{dN}^{'}.		\cos(\theta)\}.\sin(\phi) + \textrm{dN}^{''}.\cos(\phi)\\
		\textrm{d}r^{'} + \dfrac{A-C}{AC}.p^{'}q^{'}\textrm{dt}&= -\{\textrm{dN}.\cos(\theta)+ \textrm{dN}^{'}.\sin(\theta)\}.\cos(\phi) + \textrm{dN}^{''}.\sin(\phi)\\
	\end{aligned}
	\label{eq:03a}
\end{equation}

	On the left side of (\ref{eq:03a}), (\textit{p,q,r}) stand for the Euler angles ($\omega_1$, $\omega_2$ and $\omega_3$) of equations (\ref{eq:01}) and (\textit{p',q’,r’}) = (\textit{Cp, Aq, Br}), still with (\textit{A,B,C}) being the Earth’s moments of inertia. Let (\textit{x, y, z}) be the coordinates of the Earth’s center of gravity, (\textit{x’,y’,z’}) the coordinates of a mass element $dm$ of Earth. Then:
\begin{equation}
	\begin{aligned}
		\int (\dfrac{x^{'}dy^{'} - y^{'}dx^{'} }{dt}). dm &= N \\
		\int (\dfrac{x^{'}dz^{'} - z^{'}dx^{'} }{dt}). dm &= N^{'} \\
		\int (\dfrac{y^{'}dz^{'} - z^{'}dy^{'} }{dt}). dm &= N^{''} 
	\end{aligned}
	\label{eq:03b}
\end{equation}

are classical expressions for the second order inertia tensor. $\theta$ and $\psi$  are as defined is \fig \ref{Fig:01}. How can \textit{N, N’} and\textit{ N”} be evaluated without knowing the motions of all particles $dm$ in and out of the Earth? In a bold move, \citet{Laplace1799}, volume 1, book 5, page 305, paragraph 3, assumes that \textit{dN}, \textit{dN’} and \textit{dN”} can be computed from the positions and motions of the celestial bodies that act on it.\\

Laplace starts with the Sun, its mass \textit{L} and distance \textit{r$_1$}. (\ref{eq:03a}) becomes:
\begin{equation}
	\begin{aligned}
		dp + \dfrac{B-A}{C}.q.r.dt &= \dfrac{3L}{r_1^5}.(\dfrac{B-A}{C})x.y\\
		dq + \dfrac{C-B}{A}.r.p.dt &= \dfrac{3L}{r_1^5}.(\dfrac{C-B}{A})y.z\\
		dr + \dfrac{A-C}{B}.p.q.dt &= \dfrac{3L}{r_1^5}.(\dfrac{A-C}{B})x.z\\
	\end{aligned}
	\label{eq:03c}
\end{equation}

This system is close to that proposed by \citet{Guinot1976}, page 530, system 1:
\begin{equation}
	\begin{aligned}
		A \dot{p} + (C-A)qr & = L_t\\
		B \dot{q} + (A-C)rp & = M_t\\
		C \dot{r} + (B-A)pq & = N_t\\
	\end{aligned}
	\label{eq:03d}
\end{equation}
\end{subequations}
Changing the directions of the axes and since \textit{B} $\sim$ \textit{A}, (\ref{eq:03c}) and (\ref{eq:03d}) are equivalent. But for Laplace, all terms on the right side are celestial (astronomical), whereas for Guinot the torques $L_t$,  $M_t$ and  $N_t$ (not to be mistaken for \textit{N} in \ref{eq:03a}) can be external or internal to the Earth. For the Laplace formulation, one must take into account all planets that can produce effects one wants to account for. For instance, Laplace gives the full equations with the Sun and Moon included:
\begin{subequations}
\begin{equation}
\theta = h + \dfrac{3m}{4n}.(\dfrac{2C-A-B}{C}) 
\left\{
 	\begin{array}{lll}
		    &\dfrac{1}{2}.\sin(\theta).[\cos(2\nu) + \dfrac{\lambda m}{m'}\cos(2\nu')] \\
        	& -(1+\lambda).m.\cos(\theta).\Sigma \dfrac{c}{f}.\cos(ft+\varsigma) \\
        	& + \dfrac{\lambda c'}{f'}.\cos(\theta).\cos(f' t + \varsigma')
 	\end{array}
\right.	
\label{eq:04a}
\end{equation}
\begin{equation}
 \begin{array}{ll}
\dfrac{d \psi}{dt} &=  \dfrac{3m}{4n}.(\dfrac{2C-A-B}{C}) * \\
&\left\{ 
 \begin{array}{lll}
        & (1+\lambda).m.\cos(\theta) - \dfrac{\cos(\theta)}{2}.\dfrac{d}{dt}[\sin(2\nu) + \dfrac{\lambda m}{m'}.\sin(2\nu')]\\
        &  (1+\lambda).m.\dfrac{\cos^2(\theta)-\sin^2(\theta)}{\sin(\theta)}.\Sigma c.\cos(ft+\varsigma)\\
        &  \lambda .m\dfrac{\cos^2(\theta)-\sin^2(\theta)}{\sin(\theta)}. c'.\cos(f't+\varsigma')
    \end{array}
\right.
 \end{array}
\label{eq:04b}
\end{equation}
\end{subequations}

The inclination $\theta$ of the rotation axis has the current value \textit{h} in (\ref{eq:04a}). $\dfrac{d\psi}{dt}$ is linked to the Earth’s rotation, therefore to the lod. On the right side of (\ref{eq:04a} and \ref{eq:04b}) are the ephemerids and masses of the Moon and Sun that enter the classical theory of gravitation (see Appendix A in \citet{Lopes2021} for more details). Length of day and polar inclination are clearly connected by equations (\ref{eq:04a} and \ref{eq:04b}). Thus, Laplace reduces the problem to a system of two equations for the inclination and time derivative of the declination of the Earth’s rotation axis. $\theta$ and $\dfrac{d\psi}{dt}$ (and the norm that can be considered as a known constant) give the direction of the polar rotation axis and its variations. The time difference (in ms) between the theoretical and measured Earth rotation is proportional to $\dfrac{\psi}{v}$, $v$ being the rotation velocity (and the Earth’s radius is a constant). Either $\psi$ alone, or $v$ alone, or both can vary. We assume the former, since the mean rotation rate apparently remains constant, as was already noted above, and equation (\ref{eq:04b}) implies studying the time derivative of declination of the rotation axis, thus studying a quantity that is linearly related to the derivative of lod.

\section{Confronting the theory with the observations}
\begin{figure}
\centerline{\includegraphics[width=\columnwidth]{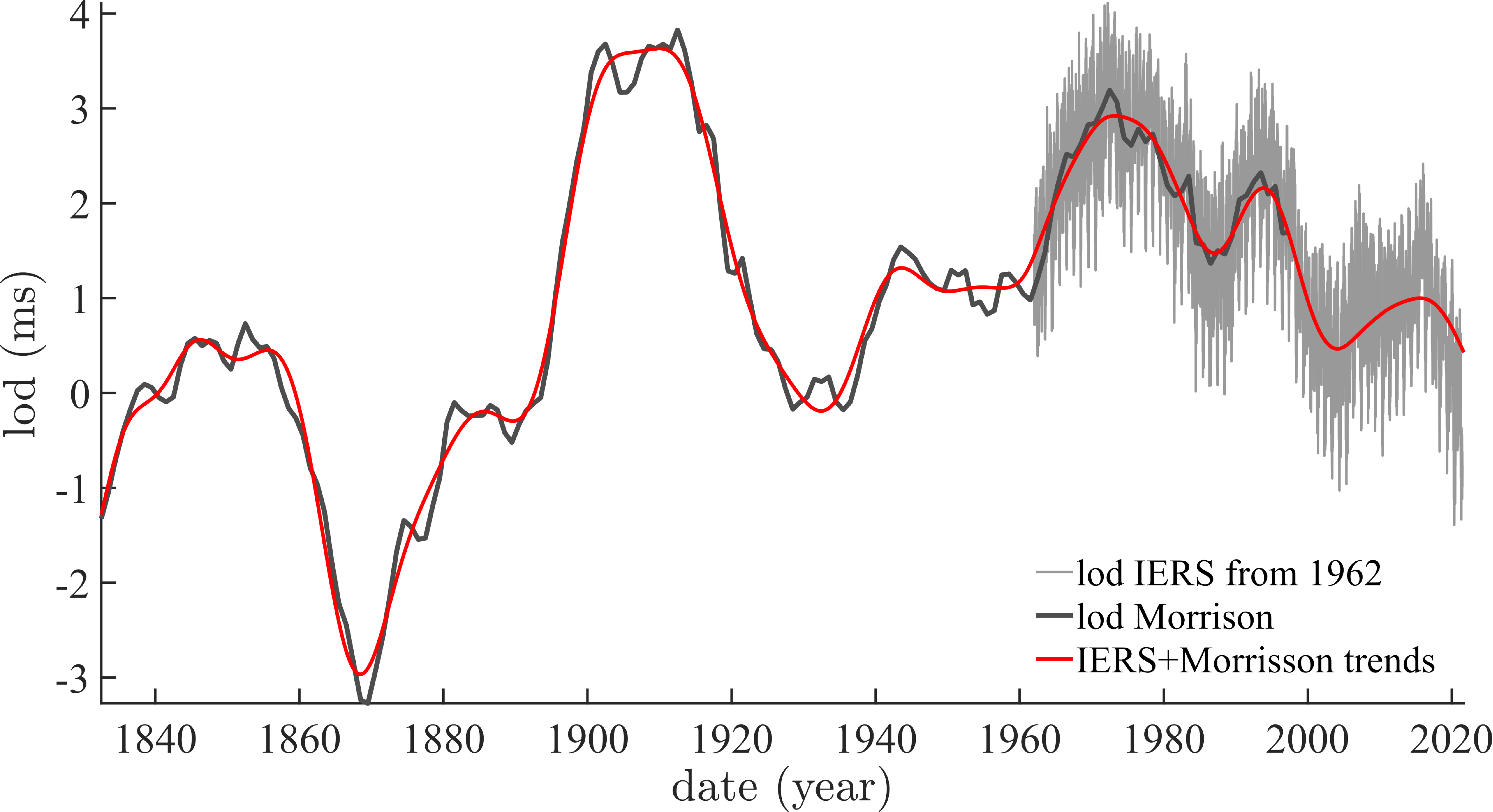}} 	
\caption{Monthly values of length of day data (LUNAR97, 1832-1997; bold black curve) from \citet{Gross2001} and daily values (1962-Present, gray curve) from \textbf{IERS}. Superimposed are their}
\label{Fig:02}
\end{figure}

\citet{Stephenson1984} have compiled the reference data for lod from 700AD onwards. This has allowed \citet{Gross2001} to build a monthly data set of lod from 1832 to 1997 (LUNAR97). We have combined it with the daily data provided by \textbf{IERS} (\textit{International Earth Rotation Service}) since 1962\footnote{https://www.iers.org/IERS/EN/DataProducts/data.html}. \fig (\ref{Fig:02}) shows the two data sets, and their (smoother) trends (the red curve). In order to determine these trends, we have applied Singular Spectrum Analysis (\textbf{SSA}; see \citet{Golyandina2013}; \citet{Lemmerling2001}; \citet{Golub1971}). The trends are the first, leading (in terms of pseudo-period and amplitude) components of the data series. The trends of the two series are smoothly continuous where they meet (1962).\\

\begin{figure}
	\begin{subfigure}[b]{\columnwidth}
		\centerline{\includegraphics[width=\columnwidth]{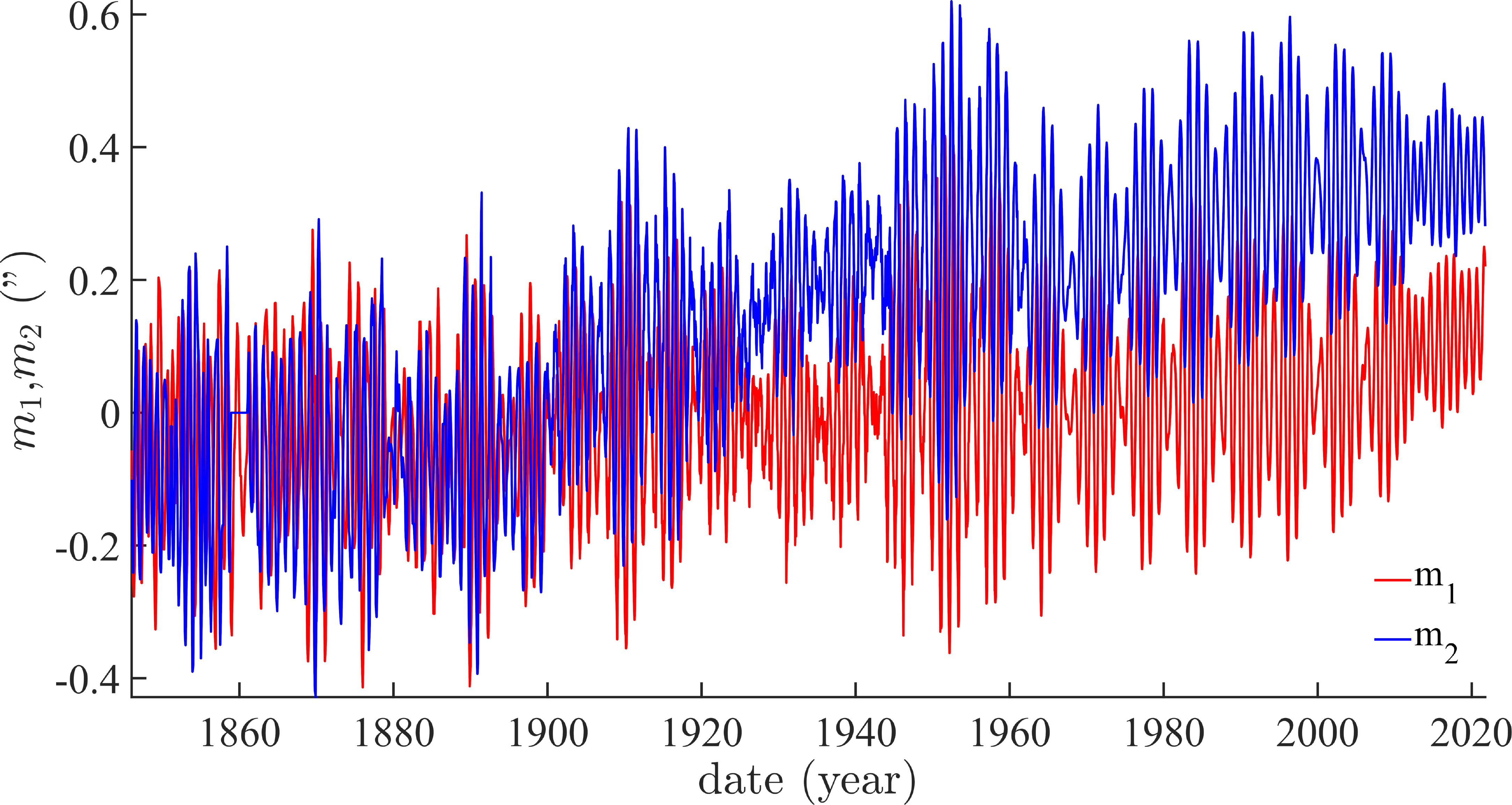}} 	
		\caption{The $m_1$ and $m_2$ components of polar motion from \textbf{IERS} (from 1846 to the Present).}
		\label{Fig:03a}
	\end{subfigure}
	\begin{subfigure}[b]{\columnwidth}
		\centerline{\includegraphics[width=\columnwidth]{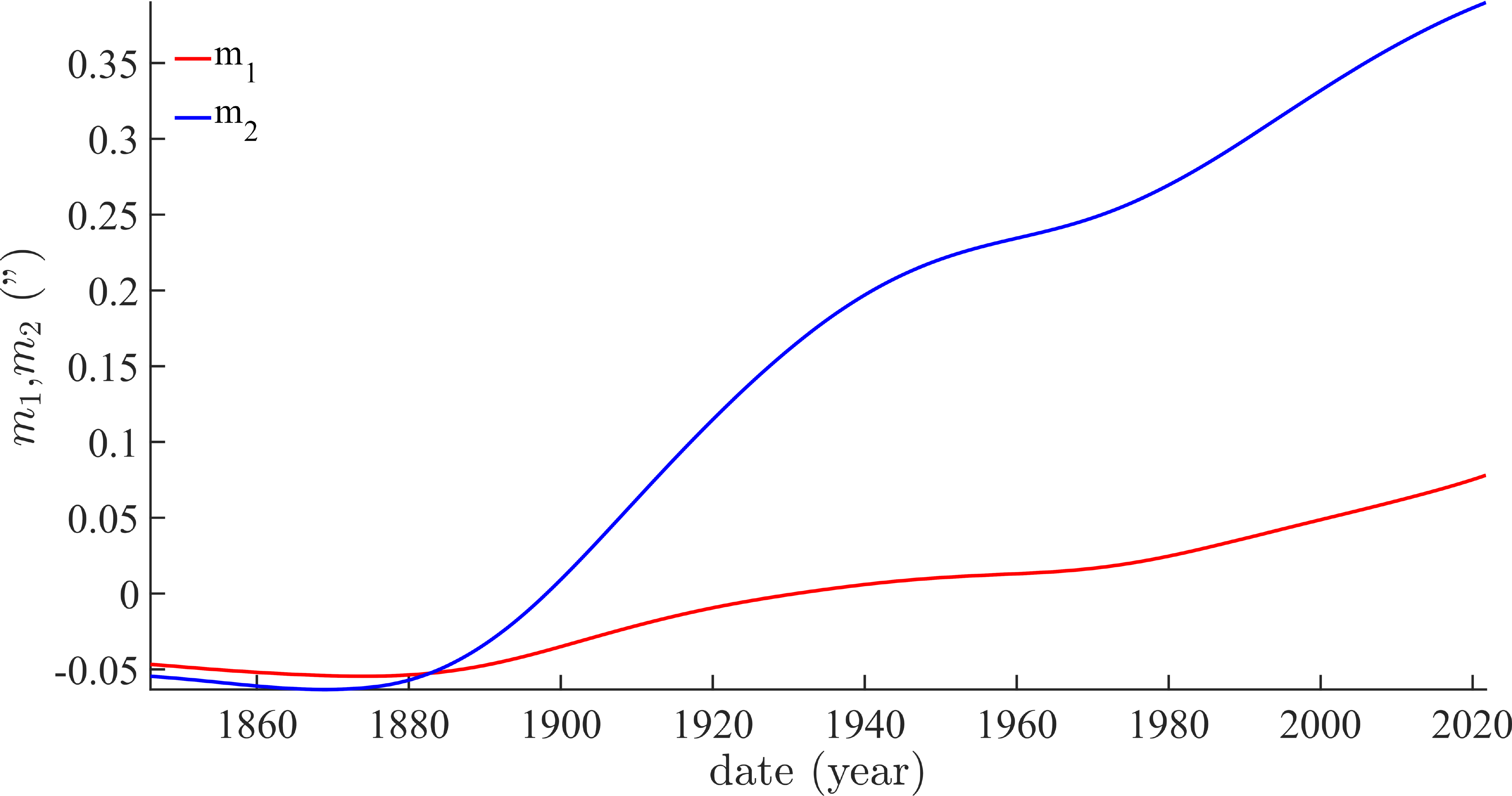}} 	
		\caption{The trends of components $m_1$ and $m_2$ from 1846 to the Present, extracted using \textbf{SSA}.}
		\label{Fig:03b}
	\end{subfigure}
	\caption{Polar motion from \textbf{IERS}}
\end{figure}

\begin{figure}[h]
	\begin{subfigure}[b]{\columnwidth}
		\centerline{\includegraphics[width=\columnwidth]{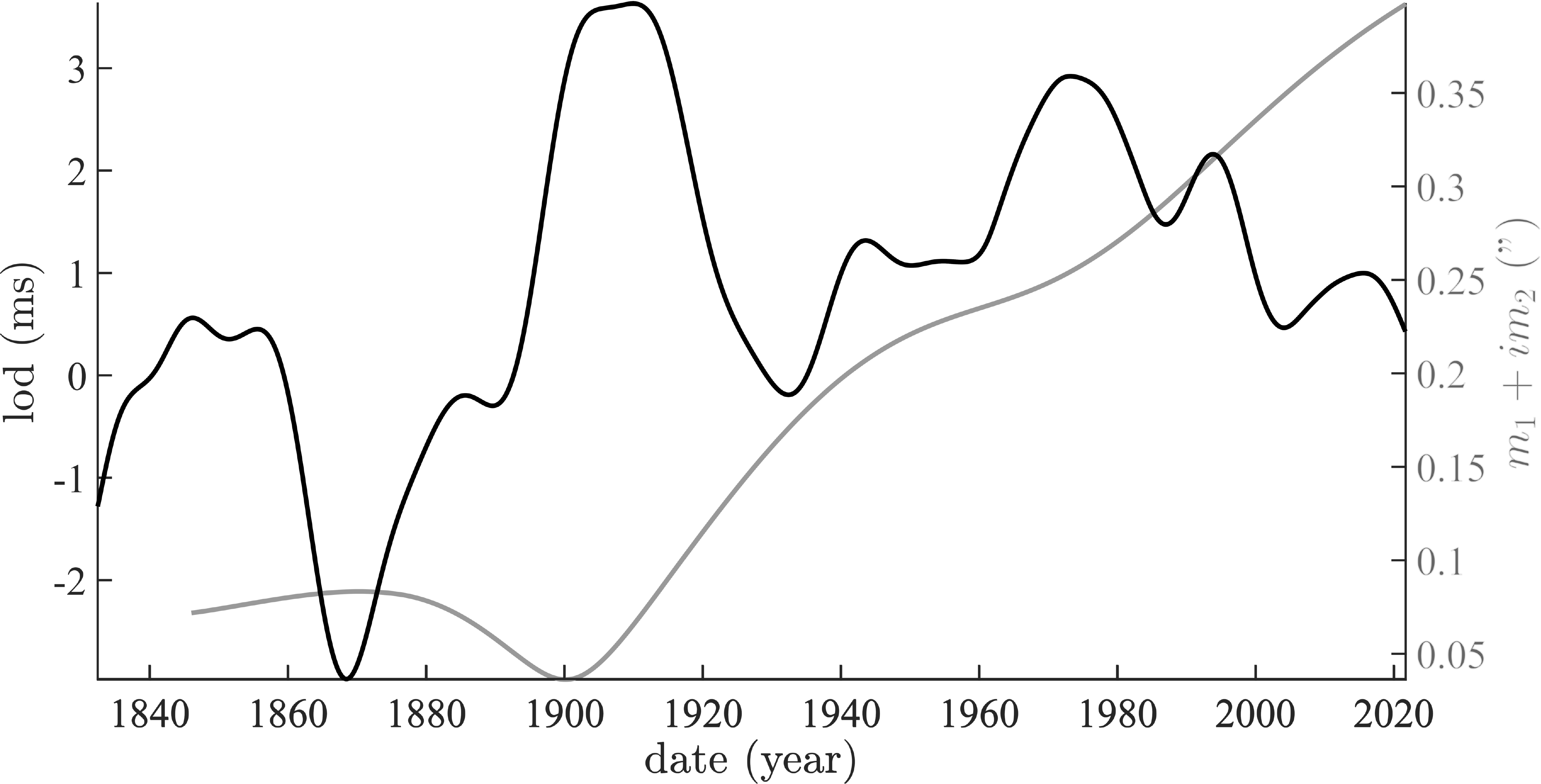}} 	
		\caption{Comparison of the trends of the Morisson/\textbf{IERS} lod (black) and of polar motion \textbf{m} (gray)}
		\label{Fig:04a}
	\end{subfigure}
	\begin{subfigure}[b]{\columnwidth}
		\centerline{\includegraphics[width=\columnwidth]{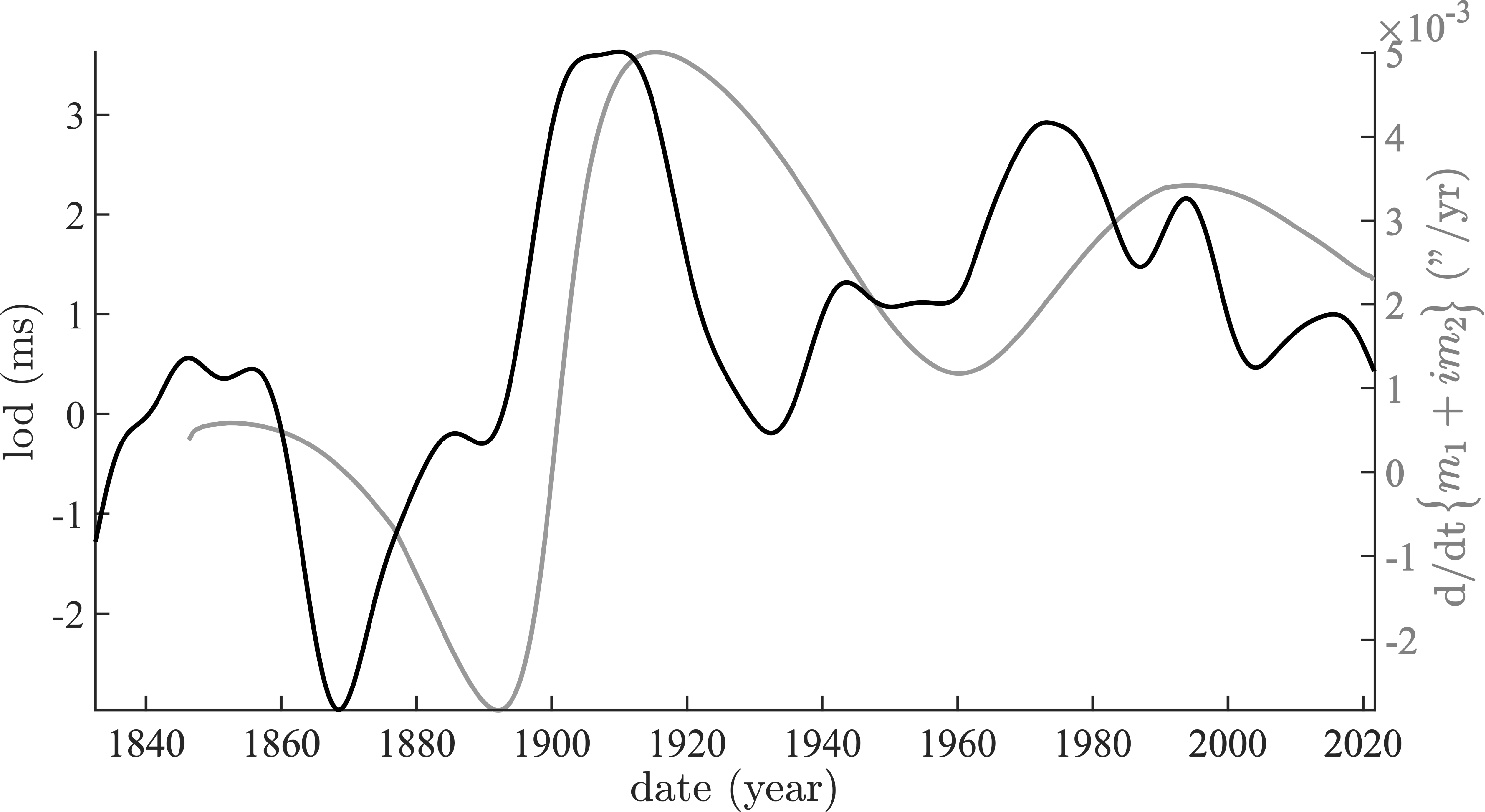}} 	
		\caption{Comparison of the trend of the Morisson/\textbf{IERS} lod (black) and of the derivative of the Markowitz drift (gray).}
		\label{Fig:04b}
	\end{subfigure}
	\begin{subfigure}[b]{\columnwidth}
		\centerline{\includegraphics[width=\columnwidth]{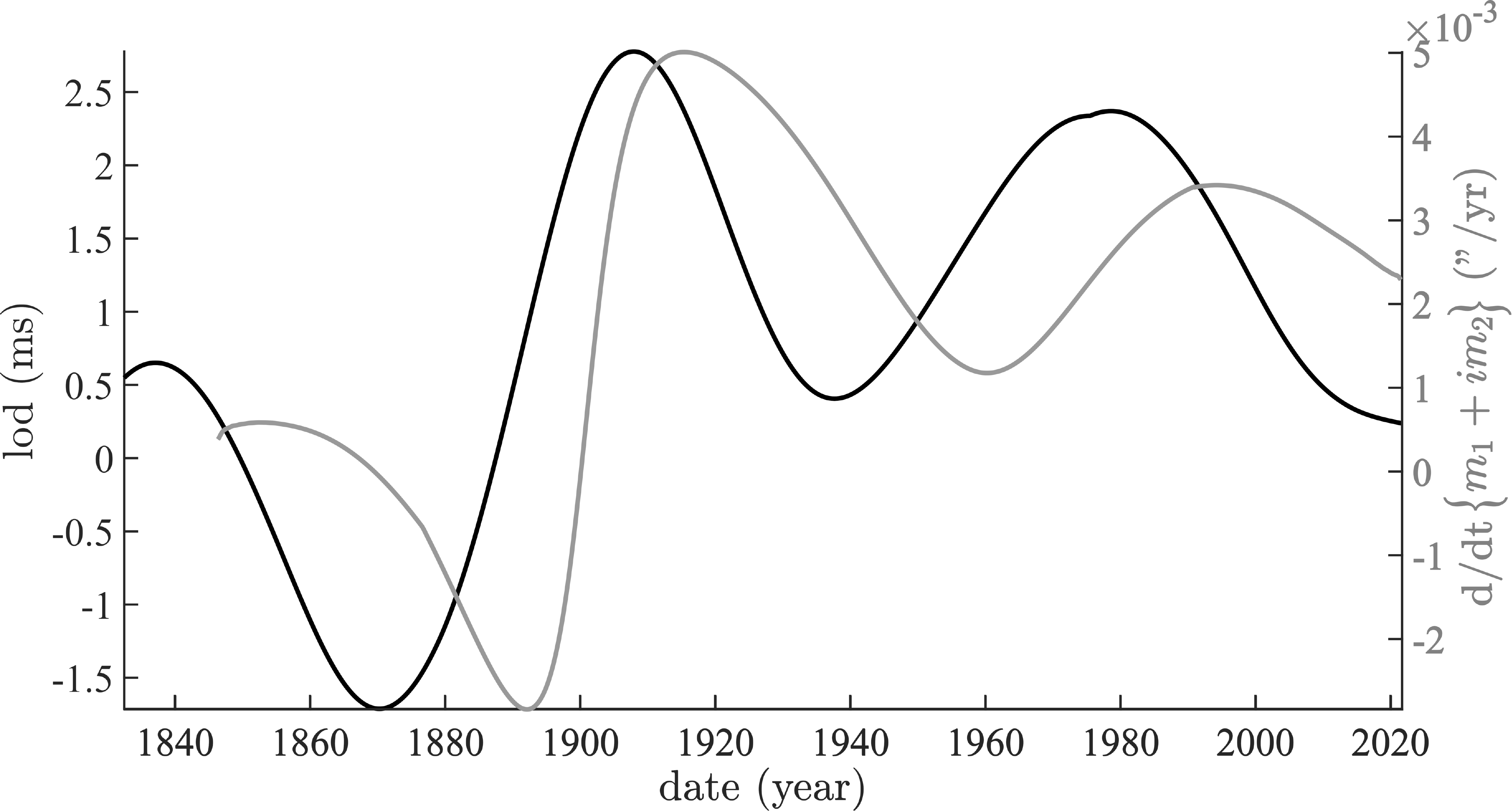}} 	
		\caption{Comparison of the smoothed trend of the Morisson/\textbf{IERS} lod (black) and of the derivative of the Markowitz drift (gray).}
		\label{Fig:04c}
	\end{subfigure}
		\caption{Polar motion from \textbf{IERS}}
\end{figure}

The data for the $ m_1$ and $m_2$ components of polar motion since 1846 are also available from \textbf{IERS} and are shown in  \fig  \ref{Fig:03a}. Their respective trends, extracted by \textbf{SSA}, are shown in \fig \ref{Fig:03b}. $m_1$ and $m_2$ are used to compute a global trend of \textbf{m} ($= m_1 + im_2$), called the Markowitz drift (\citet{Markowitz1968}). This is displayed in  \fig  \ref{Fig:04a} as a thinner gray curve, and compared to the Morrison/IERS trend of lod (thicker black curve). These curves are in excellent agreement with previous determinations (by \textit{e.g.} \citet{Stoyko1968}; \citet{Hulot1996}), though they are smoother due to \textbf{SSA} extraction. We recall that the Markowitz drift is one of the three main components of polar motion along with the Chandler free oscillation and the forced annual oscillation (\textit{e.g.} \citet{Gibert1998}; \citet{Zotov2012}; \citet{Lopes2021}). The magnitude of the Markowitz drift is on the same order as plate tectonic velocities, that originally made it quite difficult to detect.\\

Equations (\ref{eq:04a}) and (\ref{eq:04b}) imply an integrative link between  $\theta$ and$\dfrac{d\psi}{dt}$ , that is polar motion and length of day. On \fig \ref{Fig:04b}, we show again the lod curve pictured in \fig \ref{Fig:04a} (thicker black curve) and the derivative of the Markowitz drift (thinner gray curve). Despite extraction by the \textbf{SSA} method, the lod curve is still affected by higher frequency variations. This may be due to the fact that the sampling rate jumps from monthly to daily in 1962 and to the presence of a derivative (that is a high-pass filter). We smooth further (moving average windows of 10 years) the lod curve (thicker black curve; \fig \ref{Fig:04c}) that is displayed with the derivative of the Markowitz drift already shown in \fig \ref{Fig:04b} (thinner gray curve). \\

The two curves of \fig \ref{Fig:04c} are quite similar, with a lag of about a decade (lod and polar motion being in quadrature). This is consistent with the integration link between the two. We have computed the relative phase and amplitude variations that would put the two curves in best agreement (normalized to take care of the different units). We have done this by applying the simulated annealing technique (\citet{Kirkpatrick1983}) in order to bring the pole motion curve of \fig \ref{Fig:04c} (gray curve) into superposition with lod (black curve). \\

\begin{figure}
\centerline{\includegraphics[width=\columnwidth]{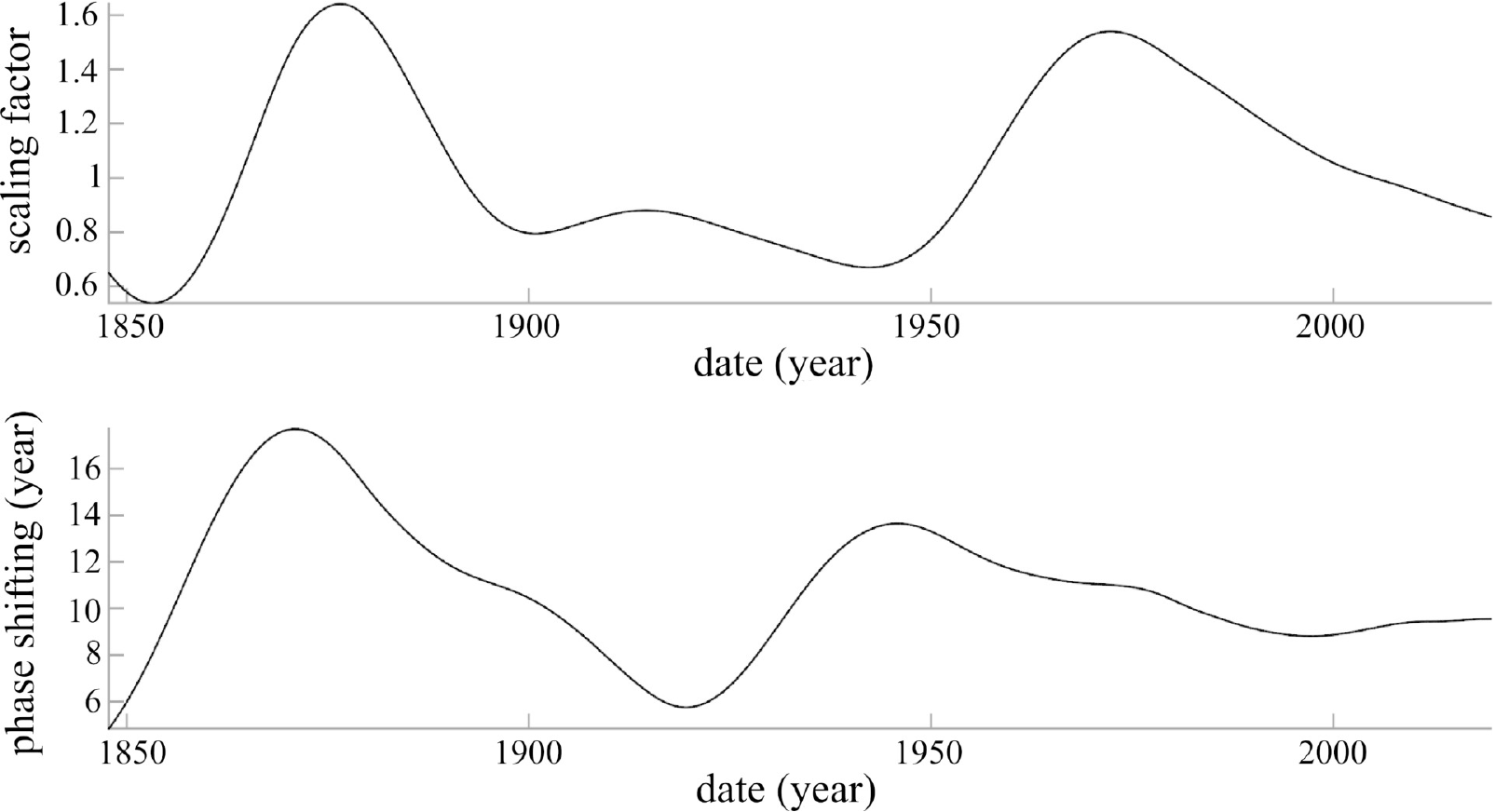}} 	
\caption{Scaling factor (top) and phase shift (bottom) of the operator that brings the two curves of Figure \ref{Fig:04c} in best agreement from 1846 to the Present.}
\label{Fig:05}
\end{figure}

 \fig \ref{Fig:05} shows the results, the amplitude and phase shift respectively at the top and bottom. We do not display the results of this inversion on waveforms since they are practically superposable (the dynamic time warping algorithm calculates the modulations of phase and amplitude of the operator that transforms the black curve of \fig \ref{Fig:04c} into the gray curve; the end result is two curves that are identical to 99\%). The scaling factor (or amplitude of the operator) required to transform the normalized polar motion to the normalized lod ranges between 0.6 and 1.6 and averages 1.03 $\pm$ 0.30: to first order, the amplitudes of the two geophysical quantities \textbf{m} and lod evolve in parallel. The larger differences occur between 1870 and 1890, and a century later between 1970 and 1990 (\fig \ref{Fig:05} top).  The largest “jump” in both lod (4 ms) and pole motion (8.10-3 ”/yr) takes place between 1870 and 1900 (\fig \ref{Fig:04c}). The phase shift of that operator ranges between 6 and 16 years and averages 10.7 $\pm$ 3.0 yr. Note that during the period between 1920 and 1940, when the amplitude factor is close to a minimum ($\sim$ 0.7), the Chandler free oscillation of polar motion suffers a well-known phase jump of $\pi$. 

\section{Discussion and concluding remarks}
Passing from the “classical” system of equations (\ref{eq:04a} and \ref{eq:04b}, \citet{Laplace1799}) to the “modern” system of equations (\ref{eq:02}, \textit{e.g.} \citet{Lambeck2005}) is not complicated but is rather lengthy. There is no contradiction between the two formulations. However, each can be read with its own emphasis. The main difference between what Laplace and Poincaré write, and what Lambeck and Guinot write arises at the step of system (\ref{eq:03a}), leading either to (\ref{eq:02}) or to (\ref{Fig:04a} and \ref{Fig:04b}). How does one account for system (\ref{eq:03b}), that is for the distribution and motions of masses inside or outside Earth, at any place and instant? Laplace solves the point with a hypothesis: if one does not have access to masses and accelerations, one can still determine the force budget. For Laplace, these forces are exclusively external, leading to system (\ref{eq:03c}), and then (\ref{eq:04a} and \ref{eq:04b}) with the Sun and Moon included. As a consequence, internal and external mass motions only serve to dissipate the energy received by polar motion from all celestial bodies (see \citet{Lopes2021}, Appendix 1). With Lambeck and Guinot’s formulation (\ref{eq:02}), one has to resort to purely terrestrial forces to explain mantle motions and changes in climate. Emphasis is on the separation between the polar coordinates (polar motion) and the third coordinate (linked to lod), and on determining the excitation functions, that can be external or internal to the Earth. These are mathematical functions that “\textit{include all factors that perturb the rotation motion}” (\citet{Lambeck2005}, page 36). These excitation functions are listed by \citet{Lambeck2005}, page 47: “\textit{the excitation functions consist of contributions from (i) redistribution of mass, (ii) relative motion of matter, and (iii) torques}”, the difference between (i) and (ii) being tenuous. For Guinot (system \ref{eq:03d}) the torques can be one or more external and internal actions. \\

In short, in the “modern” presentation, polar motion and length of day are decoupled and correspond to different physical mechanisms, whereas in the “classical” presentation polar motion involves coupling of the inclination of the rotation pole and the derivative of its declination. \\

In a previous \textbf{SSA} analysis of lod using more than 50 years of \textbf{IERS} observations, \citet{LeMouel2019a} find nine components, all likely astronomical – thus external to the Earth (“trend”, $\sim$ 80 yr, 18.6 yr, 11 yr, 1 year, 0.5 yr, 27.54 days, 13.66 days, 13.63 days, 9.13 days). The QBO at 2.36 yr is interpreted as a Sun-related oscillation. The lunar components at 13.63 and 13.66 days could contain a solar contribution. The longer periods, 1 yr, 11 yr, 18.6 yr and ~80 yr (Markowitz, \fig \ref{Fig:04c}) are common to lod and polar motion. If the link between the two is as established by \citet{Laplace1799}, then the straightening of the inclination of the axis of rotation would accompany a decrease in lod. The Earth indeed straightens up (\textit{cf.} \citet{Stoyko1968} and \fig  \ref{Fig:03b}). Also, if this link is valid, all the components with extraterrestrial periods should be present in the series of sunspots; and indeed they are (\textit{e.g.} \citet{Courtillot2021}). As far as quasi annual components of sunspots are concerned, there is no exact 1 yr line, but two nearby lines with periods 0.93 and 1.05 yr (\citet{LeMouel2020}, Table 01), that could be luni-solar commensurabilities ((365.25-28)/365.25 = 0.92 and (365.25+28)/365.25 = 1.07, 28 days being the Moon’s synodic period (\textit{cf.} \citet{Courtillot2021}; \citet{Lopes2021}; \citet{Bank2022}). Can a sufficiently strong source of energy be found? Indeed it has been known for some time (\textit{e.g.} \citet{Dickman1989}; \citet{Chao1995}; \citet{Varga2005}) that the components with luni-solar periods found above account for 95\% of the total variance of the lod signal (\citet{LeMouel2019a}). \\

Another benefit of using the original Laplace approach is the determination of the period of the Euler free oscillation. In one of the first analyses of actual observations of polar motion, \textcolor{blue}{Chandler (1891a, 1891b)} discovered the wobble that now bears his name. Chandler wrote that "\textit{the general result of a preliminary discussion is to show a revolution of the earth's pole in a period of 427 days.}" The observed period of this free oscillation is much larger than the theoretical value of 306 days, even more so now that the period has reached about 433 days (\textit{e.g.} \citet{Zotov2012}; \citet{Lopes2017}). As noted at the end of section 3, the envelope of the Chandler oscillation is strongly modulated, reaching a quasi-minimum around 1930 with its well known phase jump of $\pi$. The duration and the modulation of the Chandler wobble require a source of excitation. Earth elasticity, large earthquakes (\textit{e.g.} \citet{Mansinha1967}; \citet{Dahlen1971}; \citet{OConnell1976}; \citet{Gross1986}), or external forcing by the fluid envelopes (\textit{e.g.} \citet{Rochester1984};\citet{Gross2000};\citet{Aoyama2001}; \citet{Brzezinski2002}; \citet{Desai2002}; \citet{Lambeck2005}, chapter 7) have been successively invoked. \\

In the more classical reading of Laplace, still considering an elastic (or plastic) Earth, we have seen that the equations (\ref{eq:04a}) and (\ref{eq:04b}) allow one to calculate the period of the Euler free oscillation. This oscillation actually varies with the pole inclination $\theta$ from 306 to 578 days. A transition to a double period ($\sim$ 430 and ($\sim$ 433 days:) has taken place at the time of the 1930 phase jump of the Chandler oscillation; this can be accounted for by variations in polar inclination (\textit{cf.} \fig \ref{Fig:04c}). As is the case for a top, as recalled in section 2, the various rotations (precession, etc…) vary with the inclination of the rotation axis. \\

\begin{figure}
\centerline{\includegraphics[width=\columnwidth]{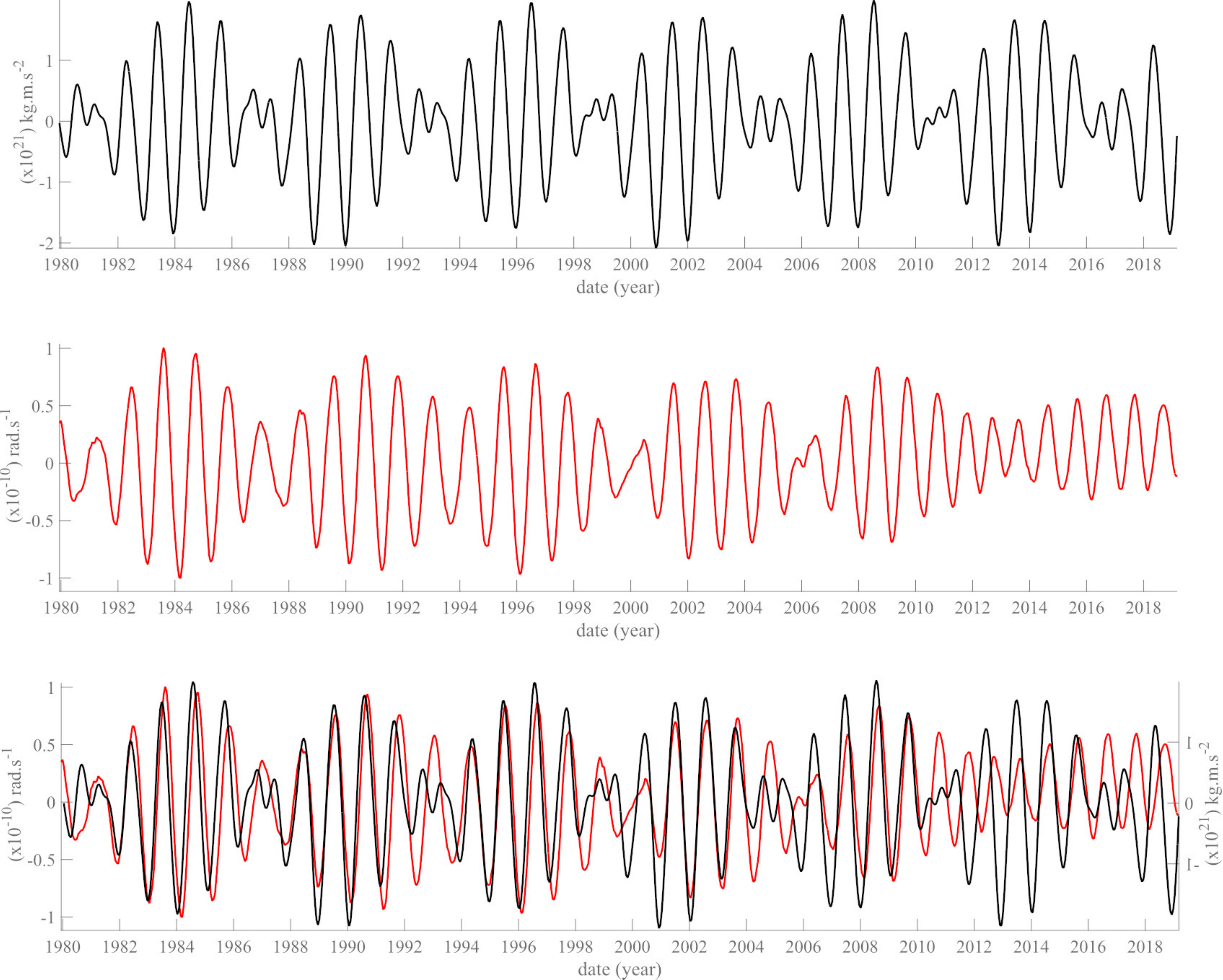}} 	
\caption{Top curve (in black): the sum of the forces of the four Jovian planets affecting Earth (ephemerids from the \textbf{IMCCE}). Middle curve (in red): the $m_1$ component of polar motion (1980–2019), reconstructed with \textbf{SSA} and with the trend (Markowitz) removed. Bottom: superposition of the 2 curves. From \citet{Lopes2021}, Figure 03.}
\label{Fig:06}
\end{figure}

\textcolor{blue}{Laplace’s (1799)} calculations and conclusions have been confirmed, prominently by \citet{Poincare1899}. Based on these equations, a link between the rotations and the torques exerted by the planets of our solar system is expected. Indeed, we have shown elsewhere the influence of Jovian planets on the Sun (sunspots, \textit{cf.} \citet{Courtillot2021}), and Earth (\citet{Lopes2021}, Figure 11). In the latter paper, we show the similarity between the envelope of the Chandler oscillation and the ephemerids of Neptune. Figure 03 of \citet{Lopes2021}, reproduced here as \fig \ref{Fig:06}, shows the remarkable agreement between the sum of forces exerted by the four Jovian planets and the $m_1$ component of polar motion. \\

As far as the length of day is concerned, we have seen in \fig \ref{Fig:02} that since 1970 it tends to decrease, \textit{i.e.} the Earth’s velocity of rotation tends to increase. The same phenomena envisioned above (earthquakes, variations in the fluid envelopes,…) are seen as potential causes of this trend (e.g \citet{Rochester1984}; \citet{Gross2004}; \citet{Landerer2007}; \citet{Chen2019}; \citet{Afroosa2021}). Also, the recent acceleration of rotation velocity, that contradicts previous models, may find a simple explanation with Laplace’s formalism (\textit{e.g.} \citet{Trofimov2021}). In closing, we emphasize that all these results are based on actual data and the validity of Laplace’s astronomical hypothesis, not on a model.

\bibliographystyle{aa}

\end{document}